\documentclass[%
 reprint,
 superscriptaddress,
 amsmath,amssymb,
 aps,
prapplied,
floatfix,
]{revtex4-2}
\usepackage{graphicx}% Include figure files
\usepackage{dcolumn}% Align table columns on decimal point
\usepackage{bm}% bold math
\usepackage{siunitx}
\usepackage{amsmath}
\usepackage{orcidlink}
\usepackage{lipsum}

\usepackage[T1]{fontenc}

\usepackage{hyperref}% add hypertext capabilities

\usepackage{algorithm}[compatible]
\usepackage{algpseudocode}

\usepackage[normalem]{ulem}
\definecolor{green}{rgb}{0.0,0.5,0.0}

\begin{document}
%%%%%%%%%%%%%%%%%%%%%%%%%%%%%%%%%%%%%%%%%%%%
\title{Towards autonomous time-calibration of large quantum-dot devices: \\Detection, real-time feedback, and noise spectroscopy}% Force line breaks with \\

\author{Anantha S. Rao\orcidlink{0000-0001-6272-0327}}
\email{anantha@umd.edu}
\affiliation{Department of Physics, University of Maryland, College Park, Maryland 20742, USA}
\affiliation{Joint Center for Quantum Information and Computer Science, University of Maryland, College Park, Maryland 20742, USA}

\author{Barnaby van Straaten\orcidlink{0009-0008-2113-8523}}
\affiliation{QuTech and Kavli Institute of Nanoscience, Delft University of Technology, P.O. Box 5046, 2600 GA Delft, The Netherlands}

\author{Valentin John}
\affiliation{QuTech and Kavli Institute of Nanoscience, Delft University of Technology, P.O. Box 5046, 2600 GA Delft, The Netherlands}

\author{Cécile X. Yu}
\affiliation{QuTech and Kavli Institute of Nanoscience, Delft University of Technology, P.O. Box 5046, 2600 GA Delft, The Netherlands}

\author{Stefan D. Oosterhout}
\affiliation{QuTech and Netherlands Organisation for Applied Scientific Research (TNO), Delft, The Netherlands}

\author{Lucas Stehouwer}
\affiliation{QuTech and Kavli Institute of Nanoscience, Delft University of Technology, P.O. Box 5046, 2600 GA Delft, The Netherlands}

\author{Giordano Scappucci\orcidlink{0000-0003-2512-0079}}
\affiliation{QuTech and Kavli Institute of Nanoscience, Delft University of Technology, P.O. Box 5046, 2600 GA Delft, The Netherlands}

\author{M. D. Stewart,~Jr.}
\affiliation{National Institute of Standards and Technology, Gaithersburg, Maryland 20899, USA}

\author{Menno Veldhorst\orcidlink{0000-0001-9730-3523}}
\affiliation{QuTech and Kavli Institute of Nanoscience, Delft University of Technology, P.O. Box 5046, 2600 GA Delft, The Netherlands}

\author{Francesco Borsoi\orcidlink{0000-0001-9398-7614}}
\affiliation{QuTech and Kavli Institute of Nanoscience, Delft University of Technology, P.O. Box 5046, 2600 GA Delft, The Netherlands}
\affiliation{NNF Quantum Computing Programme, Niels Bohr Institute, University of Copenhagen, Blegdamsvej 17, 2100 Copenhagen, Denmark}

\author{Justyna P. Zwolak\orcidlink{0000-0002-2286-3208}}
\email{jpzwolak@nist.gov}
\affiliation{Department of Physics, University of Maryland, College Park, Maryland 20742, USA}
\affiliation{Joint Center for Quantum Information and Computer Science, University of Maryland, College Park, Maryland 20742, USA}
\affiliation{National Institute of Standards and Technology, Gaithersburg, Maryland 20899, USA}

\date{\today}
%%%%%%%%%%%%%%%%%%%%%%%%%%%%%%%%%%%%%%%%%%%%%%%%%%%%%%%%%%%%%
\begin{abstract}

The performance and scalability of semiconductor quantum-dot (QD) qubits are limited by electrostatic drift and charge noise that shift operating points and destabilize qubit parameters.
As systems expand to large one- and two-dimensional arrays, manual recalibration becomes impractical, creating a need for autonomous stabilization frameworks.
Here, we introduce a method that uses the full network of charge-transition lines in repeatedly acquired double–quantum-dot charge stability diagrams (CSDs) as a multidimensional probe of the local electrostatic environment.
By accurately tracking the motion of selected transitions in time, we detect voltage drifts, identify abrupt charge reconfigurations, and apply compensating updates to maintain stable operating conditions.
We demonstrate our approach on a 10-QD device, showing robust stabilization and real-time diagnostic access to dot-specific noise processes.
The high acquisition rate of radio-frequency reflectometry CSD measurements also enables time-domain noise spectroscopy, allowing the extraction of noise power spectral densities, the identification of two-level fluctuators, and the analysis of spatial noise correlations across the array.
From our analysis, we find that the background noise at 100~$\mu$\si{\hertz} is dominated by drift with a power law of $1/f^2$, accompanied by a few dominant two-level fluctuators and an average linear correlation length of $(188 \pm 38)$~\si{\nano\meter} in the device.
These capabilities form the basis of a scalable, autonomous calibration and characterization module for QD-based quantum processors, providing essential feedback for long-duration, high-fidelity qubit operations.

\end{abstract}

\maketitle
%%%%%%%%%%%%%%%%%%%%%%%%%%%%%%%%%%%%%%%%%%%%%%%%%%%%%%%%%%%%%
\section{Introduction}
%%%%%%%%%%%%%%%%%%%%%%%%%%%%%%%%%%%%%%%%%%%%%%%%%%%%%%%%%%%%%
Semiconductor quantum dots (QDs) are a leading platform for scalable quantum processors, offering high-density integration while drawing on mature semiconductor fabrication~\cite{Vandersypen19-QCS, Burkard21-SSQ, george2024-TSQ, steinacker2025-ICS}.
Recent advances in gate-defined architectures have enabled extended one- and two-dimensional (1D and 2D) arrays with tunable interdot coupling, high-fidelity control, and scalable readout~\cite{Borsoi22-QCA, John24-TAL, Unseld2025-BCS}. %tadokoro2021-DTD, 
Yet, scaling from a few well-behaved qubits to large, interconnected registers remains limited by device variability and, critically, by temporal instabilities that require frequent recalibration.

One of the central challenges in semiconductor qubit operation is the sensitivity of each QD to its local electrostatic environment.
In planar silicon electron or germanium hole spin-qubit devices, charge noise and magnetic noise are often the dominant sources of decoherence~\cite{Dutta1981-LFF, kuhlmann2013-CSN, Connors2019-LFN, Stehouwer24-ESG}.
While isotopic purification can reduce magnetic noise, spin qubits remain susceptible to charge noise~\cite{scappucci2021-GQI}.

Because key Hamiltonian parameters---including detuning, tunnel couplings, and exchange interactions---depend sharply on the electrostatic potential, uncompensated electrostatic drift translates into operating-point detuning, loss of control fidelity, and increased dephasing~\cite{kuhlmann2013-CSN, rudolph2019-LTM, connors2022-CNS, Mehmandoost2024-DSB, capannelli2025-TSQ}.
Although certain operating regimes reduce sensitivity to charge noise (e.g., symmetry points)~\cite{Reed2016-RSC, hendrickx2024-SSO}, the locations of these operating points themselves drift in time. 
As arrays grow, manual retuning becomes impractical, motivating autonomous methods that can both \emph{stabilize} operating points and \emph{diagnose} the underlying noise processes that perturb them.

Early automation efforts have made substantial progress in device bootstrapping and coarse tuning, using signal processing, machine learning (ML), and computer vision to navigate high-dimensional gate spaces and identify charge states and relevant parameters~\cite{Hsiao20-EOT, Mills19-CAT, Zwolak21-AAQ, Rao24-MAViS}.
Once tuned, however, maintaining a multi-qubit operating point requires continuous monitoring and feedback~\cite{berritta2024-RTC}. 
Existing stabilization approaches often lock to a single sensor signal or a single transition edge~\cite{Nakajima2021-RTF}, or optimize a single sensor at optimal operating points by fitting a signal to a charge addition line~\cite{Hsiao2024-ETG}, or infer drift indirectly through device-agnostic tomography methods~\cite{Proctor2020-DTD, Gullans2024-CGC}. 

In dense arrays, these approaches face two limitations: (i) a single sensor provides only a narrow probe of a multidimensional electrostatic landscape, and (ii) sparse measurements make it difficult to disentangle global drifts from local instabilities and capacitive cross-talk. 
Moreover, as QDs are tuned away from reservoirs and become weakly coupled to sensors, robust diagnostics for \textit{core} dots become increasingly scarce.

Charge stability diagrams (CSDs) provide a natural measurement for addressing these challenges. 
A DQD charge state is defined by the (physical or virtual) plunger gates that set the DQD chemical potentials; while cross-capacitances can be compensated through virtual gates, maintaining the desired charge configuration over time is essential for long-duration operation.
During extended idling and repeated control cycles---for example, stabilizer measurements in quantum error correction---environmental fluctuations and residual cross-couplings shift the position of the target charge cell within the CSD. 
Repeated acquisition of CSDs converts this motion into a direct time series of effective gate-voltage fluctuations. 
Because charge-transition positions can shift by millivolts, commonly attributed to charging events of individual defects, tracking their trajectories provides a dual capability: it enables on-the-fly corrections of drifts and jumps and simultaneously yields quantitative access to the noise processes that drive them.

Here, we address these challenges by treating the complete set of charge transitions defining a target DQD charge state within a CSD as a multidimensional probe of the local electrostatic environment. 
We propose and validate a \emph{tracking, electrostatic recalibration, and noise spectroscopy} (TERNS) system that enables autonomous drift detection and compensation. 
TERNS leverages the full charge-transition network to geometrically characterize and track the evolution of a charge-state cell over time. 
By acquiring CSDs at regular intervals and tracking shifts of the charge cell, TERNS detects drifts away from calibrated operating points that directly encode effective gate-voltage fluctuations and can be used to compute compensating gate updates.

%%%%%%%%%%%%%%%%%%%%%%%%%%%%%%%%%%%%%%%%%%%%%
\begin{figure}
    \centering
    \includegraphics[width=\linewidth]{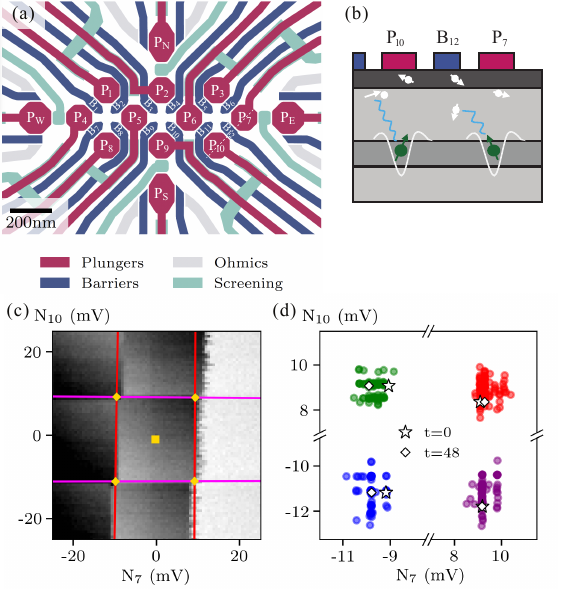}
    \caption{ 
    (a) Layout of a 3-4-3 QD device in germanium with 10 plungers and 12 barriers studied in this work. 
    From Ref.~\cite{Rao24-MAViS}.
    (b) Cartoon of a double QD system formed within a heterostructure with the electrostatic potential shown in white and a single spin qubit shown inside this potential. 
    The potential is altered by spurious interactions (green) with defects (white arrows), leading to drift in the optimal operating point (voltages, tunnel coupling, qubit frequency, etc.).
    (c) We monitor the CSD of each DQD in (2+1) space and time dimensions by extracting the center of the charge state cell and tracking its drift over time.
    (d) For each charge state cell of interest, we track for two days the locations of the four interdots at $30$~\si{\min} intervals. 
    The state at $t=0$~\si{\hour} is marked with stars and at $t=48$~\si{\hour} with diamonds.
    }
    \label{fig:intro}
\end{figure}
%%%%%%%%%%%%%%%%%%%%%%%%%%%%%%%%%%%%%%%%%%%%%

The same transition trajectories also support time- and frequency-domain spectroscopy via the Allan variance and power spectral density (PSD), respectively~\cite{Riley2008-HFS, Connors2019-LFN}. 
This resolves both two-level fluctuator (TLF) switching and slower drift.
Applying the method across multiple DQD pairs acquired simultaneously further enables spatial correlation analysis, revealing whether dominant noise sources act locally or globally across the array. 
Together, these capabilities form a foundation for automated calibration and characterization in large QD systems.

We demonstrate TERNS on a dense $3$-$4$-$3$ germanium device with ten QDs~\cite{John24-TAL, Stehouwer24-ESG}, shown in Fig.~\ref{fig:intro}(a), using long-duration repeated CSD acquisition to obtain drift trajectories for multiple DQD pairs. 
From these trajectories, we perform time- and frequency-domain noise analysis, quantify spatial correlations across the array, and extract an average correlation length of $(188\pm38)$~\si{\nano\meter}.
We further validate real-time detection of engineered shifts and quantify the sensitivity with which operating-point changes can be resolved. 
Finally, we benchmark robustness limits using simulation by systematically degrading data quality with controlled noise processes, establishing the regimes in which reliable tracking is maintained.
By unifying robust sensing of charge-state geometry with drift compensation and quantitative noise spectroscopy, TERNS provides a practical route toward the autonomous calibration and stabilization layers required for long-duration, high-fidelity operation of scalable quantum-dot qubit registers.

The remainder of the paper is organized as follows. 
Section~\ref{sec:methods} introduces the charge-cell extraction procedure and the tracking framework.
Section~\ref{ssec:tracking-results} presents long-duration measurements in the 10-QD array and the resulting drift, noise, and correlation analyses.
Section~\ref{ssec:demo-feedback} demonstrates drift detection and feedback concepts, and Sec.~\ref{ssec:noisy-sim} quantifies robustness limits using simulated data.
We conclude with a summary and outlook in Sec.~\ref{sec:conclusion}.

%%%%%%%%%%%%%%%%%%%%%%%%%%%%%%%%%%%%%%%%%%%%%%%%%%%%%%%%%%%%%
\section{Autonomous tracking and electrostatic stabilization}
\label{sec:methods}
%%%%%%%%%%%%%%%%%%%%%%%%%%%%%%%%%%%%%%%%%%%%%%%%%%%%%%%%%%%%%
Real-time acquisition and examination of charge stability diagrams (CSDs) provide a natural metric for monitoring the electrostatic environment of a QD array and for applying on-the-fly corrections to mitigate voltage drifts and charge jumps.
The core of our method is to robustly identify the coordinates of a charge-state cell within a CSD and to track its motion over time.
A charge state cell for a double-QD (DQD) is defined by the real-space hole occupations $(n_i, n_j)$ and the corresponding gate voltages applied to the plunger and barrier gates that define the chemical potentials of the DQD.
In the following sections, we provide a high-level overview of the TERNS framework.

%%%%%%%%%%%%%%%%%%%%%%%%%%%%%%%%%%%%%%%%%%%%%%%%%%%%%%%%%%%%%
\subsection{Tracking charge state cells}
\label{ssec:tracking-charge-state}
%%%%%%%%%%%%%%%%%%%%%%%%%%%%%%%%%%%%%%%%%%%%%
To track the $(n_i, n_j)$ charge state over time, we repeatedly acquire two-dimensional (2D) CSDs across two plunger gates P$_m$ and P$_n$ that define the target DQD.
To robustly extract this cell, we perform the first three stages of the modular autonomous virtualization system (MAViS) introduced in Ref.~\cite{Rao24-MAViS}, i.e., charge-sensor virtualization, plunger orthogonalization, and plunger normalization, before acquiring the sequence of CSDs.
These steps suppress stray capacitive couplings and normalize the effective transition spacing across DQDs.

The final stages of MAViS use interdot transition features from a series of CSDs to determine the cell center as a function of barrier voltage and define the barrier virtualization transformation~\cite{Rao24-MAViS}. 
This strategy becomes fragile when individual CSD must be analyzed independently, as interdot segments may be missed by the classifier or by thresholding, and performance may degrade in the presence of strong noise or latching. 
In practice, reliably identifying isolated interdot features or triple points within a single CSD is substantially more challenging than consistently detecting the dominant charge-transition lines.

To obtain a more robust estimate of the charge cell position, we leverage pixel-wise probability maps for the vertical and horizontal transitions returned by MAViS.
First, the candidate line features in the probability maps are detected using a Hough transform applied within a dynamically selected window centered on the expected location of the $(n_i, n_j)$ charge state.
The window size is adapted to ensure that the relevant transition segments are fully contained while suppressing features from neighboring charge states. 
Among the detected candidates, the two nearly parallel vertical line segments and the two nearly parallel horizontal line segments whose separation is consistent with the expected spacing between charge transitions are selected.
=
The four intersection points of these lines define the corners of a quadrilateral corresponding to the $(n_i, n_j)$ charge state cell.
The geometric center of this quadrilateral for a CSD acquired at time $t$ is,
\begin{equation}
    c(t) = \bigl(c_x(t), c_y(t)\bigr),
\end{equation}
where $c_{x/y}$ is the center coordinate along the normalized gate N$_{m/n}$.

Repeating this procedure for successive CSDs in a sequence yields a time series of charge cell centers, which we refer to as \textit{cell center trajectory},
\begin{equation}
    \mathcal{C}=\{c(t_0),c(t_1),\dots,c(t_N)\},
\end{equation}
where the time interval $T=[t_0,t_N]$ is discretized into $N\!+\!1$ steps corresponding to the measurement time stamps, and two corresponding trajectories along the N$_m$ and N$_n$ gate,  which we refer to as \textit{gate trajectories},
\begin{align}\label{eq:plunger_series}
\begin{split}
    \mathcal{C}_{m} &= \{c_x^m(t_0), c_x^m(t_1), \dots, c_x^m(t_N)\}\\
    \mathcal{C}_{n} &= \{c_y^n(t_0), c_y^n(t_1), \dots, c_y^n(t_N)\}.
\end{split}
\end{align}

The time series $\mathcal{C}$ encodes effective fluctuations of the local electrostatic potential and forms the basis for the drift, noise, and correlation analyses presented in Sec.~\ref{ssec:tracking-results}.
All subsequent estimators are computed from this cell and gate trajectories for each DQD.

%%%%%%%%%%%%%%%%%%%%%%%%%%%%%%%%%%%%%%%%%%%%%%%%%%%%%%%%%%%%%
\subsection{Drift and noise metrics}
\label{ssec:drift-metrics}
%%%%%%%%%%%%%%%%%%%%%%%%%%%%%%%%%%%%%%%%%%%%%
In the absence of intentional retuning, the electrostatic environment of a DQD evolves in time due to charge noise and, to a lesser extent, magnetic noise. 
These fluctuations shift the charge-transition network in the CSD, causing the central charge-state cell to drift slowly and, occasionally, to undergo abrupt jumps associated with charge rearrangements (e.g., TLF events).

To quantify the drift and noise, we analyze all time series $\mathcal{C}$ in both the time and frequency domains. 
Specifically, we compute (i) the Allan variance to quantify stability as a function of averaging time, (ii) the power spectral density to quantify the distribution of noise power across frequencies, and (iii) pairwise correlation coefficients between different DQDs to assess spatially correlated fluctuations across the array.
These metrics form the basis for the noise spectroscopy and correlation analyses presented in Sec.~\ref{ssec:tracking-results}.

%%%%%%%%%%%%%%%%%%%%%%%%%%%%%%%%%%%%%%%%%%%%%
\subsubsection{Allan Variance}
%%%%%%%%%%%%%%%%%%%%%%%%%%%%%%%%%%%%%%%%%%%%%
Developed originally for time and frequency standard characterization, Allan variance analysis provides a robust framework for determining the optimal averaging time in precision measurements~\cite{Riley2008-HFS}.
In the context of our work, the Allan variance serves as a diagnostic tool for the temporal stability of the QD, allowing us to identify and quantify distinct noise processes~\cite{Burnett2019-DBS, Ye2024-CIF}.

Unlike the standard variance, which often diverges for non-stationary processes, the Allan variance converges for common power-law noise types, enabling the differentiation of white noise, flicker ($1/f$) noise, and random-walk drifts.
The Allan variance $\sigma^2_A(\tau)$ for the charge cell center trajectory $\mathcal{C}=\bigoplus_i \mathcal{C}_i(\tau)$, where each $\mathcal{C}_i(\tau)$ is a disjoint sub-trajectory of $\mathcal{C}$ over a time interval of duration $\tau$, is given by
\begin{align}
    \sigma^2_A(\tau) = \frac{1}{2(\ell-1)} \sum\nolimits_{i=0}^{\ell} \left(\bar{\mathcal{C}}_{i+1}(\tau) - \bar{\mathcal{C}}_i(\tau) \right)^2, 
\end{align}
where $\bar{\mathcal{C}}_i$ denotes the mean value of a sub-trajectory $\mathcal{C}_i$ and $\ell$ is the total number of disjoint sub-trajectories~\footnote{The final sub-trajectory $\mathcal{C}_\ell$ might include fewer measurements than the initial $\ell-1$ depending on the total time $T$ and $\tau$.}.
A more practical variant of the Allan variance is the overlapping Allan variance, which improves statistical performance by utilizing a larger number of data pairs, resulting in lower variance and higher resolution~\cite{Riley2008-HFS}.

A plot of $\sigma_A^2(\tau)$ versus averaging time $\tau$ on a log-log scale reveals the dominant instability mechanisms governing the system.
The slope of the curve ($\mu$) indicates the noise color: a slope of $\mu = -1$ signifies white noise, where the variance decreases with averaging; a slope of $\mu \approx 0$ indicates a $1/f$ flicker noise floor; and $\mu = 1$  slope signals the onset of a random walk~\cite{Vliet1982-TSP}.

This analysis yields two critical insights. 
First, the minimum of the Allan deviation plot indicates the maximum time before low-frequency drifts begin to set in. 
Second, the appearance of significant local maxima indicates the presence of dominant TLFs.
The $\tau$ at which the maximum occurs is equal to the average switching time of the TLF~\cite{Burnett2019-DBS}.
By characterizing these regimes, we can isolate periods of stability from intervals dominated by drift.
We use the \textsc{AllanTools} package in Python for Allan variance calculations~\cite{Wallin18-AT}.

%%%%%%%%%%%%%%%%%%%%%%%%%%%%%%%%%%%%%%%%%%%%%
\subsubsection{Power spectral density}
%%%%%%%%%%%%%%%%%%%%%%%%%%%%%%%%%%%%%%%%%%%%%
Analyzing noise in the frequency domain is accomplished by calculating the power spectral density (PSD), which indicates how the noise power is distributed across frequencies~\cite {Riley2008-HFS}. 
From the change of the charge cell center position within a CSD $c(t)$ at different times $t$, we can directly estimate the noise process and PSD using the power spectrum in frequency space:
\begin{align}
    S^m_x(w) :&= \frac{\Delta t}{N} \left|\sum\nolimits_{n=0}^{N-1} c^m_x[t_n] e^{-i2\pi n \Delta t}\right|^2 \\
     &= S_o w^{-\alpha}
\end{align}
A spectrum with power law exponent $\alpha \approx 1$ is associated with a uniform log-normal distribution of two-level fluctuators affecting the QD~\cite{Dutta1981-LFF}. 
Additionally, Ref.~\cite{Mehmandoost2024-DSB} suggests that the density of TLFs coupled to a particular QD is directly proportional to $S_0$.
In the presence of drift or switching between two states, the spectrum would look similar to a Lorentzian with a power law $\alpha \approx 2$~\cite{Ye2024-CIF}.
For our calculations, we use the Welch method for PSD estimation~\cite{welch2003-FFT}.

%%%%%%%%%%%%%%%%%%%%%%%%%%%%%%%%%%%%%%%%%%%%%
\subsubsection{Linear correlation}
%%%%%%%%%%%%%%%%%%%%%%%%%%%%%%%%%%%%%%%%%%%%%
The Pearson correlation coefficient enables us to determine whether two plunger gates are linearly correlated.
Using Eq.~(\ref{eq:plunger_series}), we can define the gate trajectories $\mathcal{C}_{m}$ and $\mathcal{C}_{n}$ corresponding to plungers N$_m$ and N$_n$, respectively, and compute the average linear correlation as 
\begin{align}
\label{eqn:linear-correl}
    C({\rm N}_m, {\rm N}_n) = \sum_{t=0}^{N}\frac{(c_{x/y}^m(t) - \mu_{\mathcal{C}_{m}})(c_{x/y}^n(t) - \mu_{\mathcal{C}_{n}})}{\sigma_{\mathcal{C}_{m}} \sigma_{\mathcal{C}_{n}}}, 
\end{align}
where $(\mu_{\mathcal{C}_{m/n}}, \sigma_{\mathcal{C}_{m/n}})$ are the mean and standard deviation of the N$_m$ and N$_n$ gate trajectories. 

We expect $\mathcal{C}({\rm N}_m, {\rm N}_n)= 1$ for the ${\rm N}_m = {\rm N}_n$. 
If two gates ${\rm N}_{m/n}$ have high linear correlation, i.e, $|C({\rm N}_m, {\rm N}_n)| \geq 0.7$, we expect them to undergo similar noise processes and therefore be affected by similar background defects.
A low correlation, $|\mathcal{C}({\rm N}_m, {\rm N}_n)| \leq 0.3$ corresponds to decoupled local environments.

%%%%%%%%%%%%%%%%%%%%%%%%%%%%%%%%%%%%%%%%%%%%%%%%%%%%%%%%%%%%%
\subsection{Detection and feedback}
\label{ssec:detection-feedback}
%%%%%%%%%%%%%%%%%%%%%%%%%%%%%%%%%%%%%%%%%%%%%
To validate drift detection and feedback under controlled conditions, we use the constant-capacitance model simulator \textsc{QArray}~\cite{vanStraaten24-QAR, vanStraaten24-QAC}.
The simulator allows us to introduce prescribed time-dependent perturbations to the electrostatic potential and test the cell-tracking system's response.

We consider two classes of perturbations: (i) a linear-drift process, in which the charge-stability map translates uniformly in plunger space over time; and (ii) a stochastic jump process, in which the map undergoes discrete random shifts.
For the latter, we model the jump times using a Poisson process with mean waiting time $\gamma$, and draw jump magnitudes from a specified distribution.
Having introduced predictable and random drifts, we deploy our framework to detect the center of the charge stability map for all the cases.

Corrective feedback is implemented using a proportional controller that acts on the inferred displacement of the charge-state cell center. 
Denoting the detected center by $r_d(t)$ and the desired reference location by $r_0$, we define the \textit{center distance error}
\begin{align}\label{eq:dist_error}
    r_c(t) = \left\|r_d(t) - r_0\right\|_2.
\end{align}
When $r_c(t)$ exceeds a tolerance $\epsilon$, we apply a corrective update with feedback gain parameter $F_g$,
\begin{align}\label{eq:gain_factor}
    F(t) = F_g * \Theta(r_c(t) - \epsilon),
\end{align}
where $\Theta$ is the Heaviside step function.
The performance of drift detection and re-centering under these perturbations is presented in Sec.~\ref {ssec:demo_feedback}.

%%%%%%%%%%%%%%%%%%%%%%%%%%%%%%%%%%%%%%%%%%%%%%%%%%%%%%%%%%%%%
\subsection{Experimental setup}
\label{ssec:exp_setup}
%%%%%%%%%%%%%%%%%%%%%%%%%%%%%%%%%%%%%%%%%%%%%
The time-domain drift tracking experiments are performed on a dense 2D array of ten QDs defined in a planar germanium quantum well~\cite{Stehouwer24-ESG, John24-TAL}. 
The device is based on a low-disorder Ge/SiGe heterostructure grown on a germanium substrate, with the quantum well separated from the dielectric interface by a $55$-\si{\nano\meter} SiGe barrier~\cite{Stehouwer2023}.

\begin{figure*}
    \centering
    \includegraphics[width=.95\linewidth]{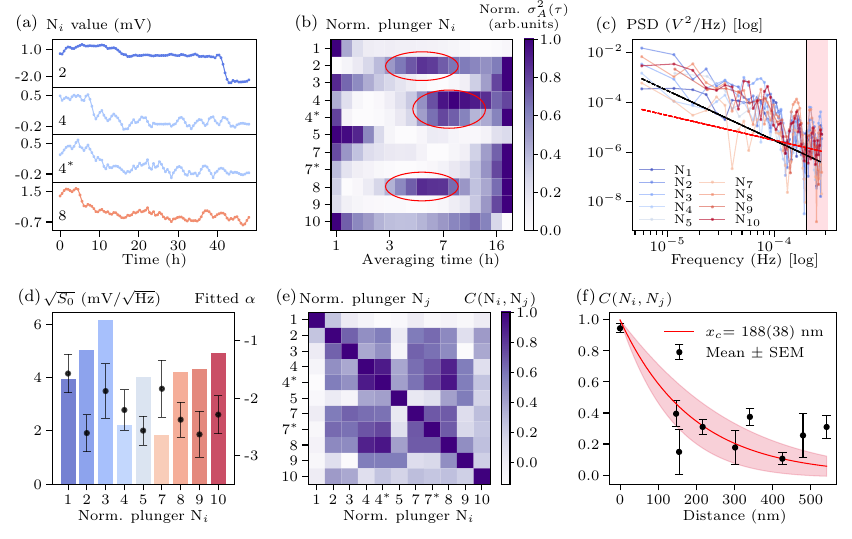}
    \caption{Time-evolution of ten QD in a planar QD device over two days. 
    (a) Extracted mean-normalized trajectory of the center of the $(n_i,n_j)$ charge state under a few selected plungers.  
    In all plunger gate voltages, we observe a gradual drift, highest in magnitude under N$_2$ where the charge stability center shifts by almost $5$~\si{\milli\volt} over two days. 
    A similar trend is observed for N$_4$ extracted from two different DQD CSD series. 
    (b) Time domain stability analysis using Allan variance with averaging time plotted in log-log scale for the same trajectory. 
    Each row corresponds to the Allan variance of a specific plunger and is normalized to be within (0,1). 
    The trajectory of the charge state under most plungers follows a power-law type noise. 
    Plunger N$_2$, N$_4$, and N$_8$ reveal a distinct peak in the center (red circles) of the Allan variance indicative of non-power law or Lorentzian-type noise, suggesting stronger coupling to a single TLF.
    (c) PSD for all QDs from the trajectory of the center of the corresponding CSD, showing that most plungers experience drift with noise that decays faster than $1/f$. 
    Black (red) dotted lines correspond to the $f^{-2}$ ($f^{-1}$). 
    (d) Extracted noise amplitude $\sqrt{S_0}$ at $100$~\textmu\si{\hertz} (colorbrs) for all ten QDs and the fitted exponents (black error bars) from the power spectral density.
    We see that the system is dominated by Red noise ($1/f^2$) at these low frequencies ($100$~\textmu\si{hertz}).
    Most plungers show a noise amplitude $(2 \text{ to } 5) \text{ mV}/\sqrt{\text{Hz}}$.
    There isn't a strictly linear trend across the plunger index, suggesting that the local environment or the specific tuning of each virtual gate affects the noise coupling differently.
    (e) The Pearson correlation coefficients (dimensionless) for all plunger pairs.
    (f) Pearson correlation coefficients as a function of distance between plungers, revealing that the correlation decays rapidly with little to no linear correlation beyond $150$~\si{\nano\meter}. 
    The solid red line represents the best-fit model, with a predicted mean correlation length of $(180\pm38)$~\si{\nano\meter}. The shaded regions indicate values within the $95~\%$ confidence interval.
    }
    \label{fig:two-day-evolution}
\end{figure*}

The device uses a multilayer gate architecture to define an array of ten QDs arranged in a 3-4-3 configuration; see Fig.~\ref{fig:two-day-evolution}(a). 
Ten plunger gates and 12 barrier gates labeled $\mathrm{P}_i$ and $\mathrm{B}_j$ for $i\in[1, \dots, 10]$ and $j\in[1, \dots, 12]$, respectively, offer control of the array’s electrostatic potential landscape.
Four additional plunger gates, labeled $\mathrm{P}_N$, $\mathrm{P}_E$, $\mathrm{P}_W$, and $\mathrm{P}_S$, control the sensor QD's potentials. 
The ten QDs are labeled D$_n$, with $n\in[1,\dots,10]$, and their four peripheral sensor QDs are labeled S$_N$, S$_E$, S$_W$, and S$_S$, based on their cardinal directions. 
The additional eight screening gates screen the electric field from the plunger gates to prevent the formation of spurious QDs. 

The array is pretuned to the few-hole regime, with each QD being either in a single-, triple-, or quintuple-hole occupancy for spin qubit manipulation~\cite{John24-TAL}.
The four charge sensors permit fast simultaneous radio-frequency- (rf)-reflectometry charge sensing in combination with video-mode acquisition and frequency multiplexing~\cite{stehlik2015, vigneau2023}.

%%%%%%%%%%%%%%%%%%%%%%%%%%%%%%%%%%%%%%%%%%%%%%%%%%%%%%%%%%%%%
\section{Results}
\label{sec:results}
%%%%%%%%%%%%%%%%%%%%%%%%%%%%%%%%%%%%%%%%%%%%%%%%%%%%%%%%%%%%%
While cross-capacitances between metallic gates can be compensated by virtualization, maintaining a fixed charge state during device operation requires monitoring and correcting slow drifts in the underlying electrostatic landscape.
The ability to access charge stability maps and infer changes in normalized plunger gate voltages in real time reveals the time dynamics of all metallic gate voltages and the noise processes affecting each normalized gate. 

In this section, we evaluate the performance of the charge-state tracking and stabilization framework when deployed on a multi-QD device and under controlled perturbations.
We first demonstrate long-duration autonomous tracking on a ten-QD array, using the extracted charge-state trajectories to characterize electrostatic drift, noise spectra, and spatial correlations. 
We then assess real-time detection and stabilization by introducing engineered voltage shifts and quantifying the accuracy, sensitivity, and robustness of the response.
Finally, we assess the performance of the tracking method in the presence of commonly observed noise.

We note that the data for plungers N$_4$ and N$_7$ are shown twice because they are used in two sets of CSDs.  
Plunger N$_6$ is not considered in the analysis because the corresponding measurements were too noisy to extract the relevant trajectory.

%%%%%%%%%%%%%%%%%%%%%%%%%%%%%%%%%%%%%%%%%%%%%
%%%%%%%%%%%%%%%%%%%%%%%%%%%%%%%%%%%%%%%%%%%%%

%%%%%%%%%%%%%%%%%%%%%%%%%%%%%%%%%%%%%%%%%%%%%%%%%%%%%%%%%%%%%
\subsection{Tracking quantum dot arrays in (2+1)D}
\label{ssec:tracking-results}
%%%%%%%%%%%%%%%%%%%%%%%%%%%%%%%%%%%%%%%%%%%%%
In a multi-QD array, it is essential to determine whether charge noise is primarily local to individual QDs or exhibits correlations across the device.
Spatially correlated fluctuations---for example, due to global gate-voltage drift or extended charge motion---can degrade the efficiency of error-correcting schemes for multi-qubit operation. 
Our automated tracking framework enables direct measurement of these spatiotemporal correlations through the time evolution of CSDs across multiple DQD pairs.

To demonstrate long-term tracking, we repeatedly acquire 2D CSDs for all target DQD pairs over 48~\si{\hour}, with approximately one scan every 30~\si{\minute}. 
Using the procedure in Sec.~\ref{ssec:tracking-charge-state}, we extract the center coordinate $c(t)$ of the $(n_i,n_j)$ charge state cell from each CSD, producing a set of drift trajectories $\mathcal{C}$ for each DQD.
These trajectories provide a direct time-domain record of electrostatic fluctuations. 
From them, we compute time- and frequency-domain stability metrics, including the overlapping Allan deviation and the PSD.

Figure~\ref{fig:two-day-evolution} summarizes the results. 
We begin with the raw center drift trajectories in Fig.~\ref{fig:two-day-evolution}(a), which reveal a combination of slow drift and discrete switching between metastable levels.
For example, the normalized plunger N$_2$, initially centered at $0$~\si{\milli\volt}, exhibits an initial drift of $1.5$~\si{\milli\volt} over the first $\sim 10$~\si{\hour}, followed by a return toward its initial value and a subsequent transition to a $-3$~\si{\milli\volt} level near $\sim 40$~\si{\hour}. 
As a consistency check, trajectories extracted for plunger N$_4$ from two different DQD-pair measurements---N$_4$-N$_8$ and N$_4$-N$_1$, shown in the two middle panels in Fig.~\ref{fig:two-day-evolution}(a)---show similar trends, supporting the robustness of the extraction procedure.
We also capture drift and switching between stable levels of plunger N$_8$, which initially increases and then continues to drift.

To quantify the stability of all gates, we plot the overlapping Allan variance in Fig.~\ref{fig:two-day-evolution}(b), normalized to $(0,1)$ for better visualization.
Trajectories exhibiting clear two-level switching---e.g., $N_2$ and $N_8$---show pronounced features in the Allan variance at characteristic averaging times, consistent with the presence of a dominant fluctuator superimposed on a slower drift.
Such a local maximum can be associated with non-power law or Lorentzian-type noise.
For the latter, the averaging time value at the peak is given by the mean switching rate between the two levels of a TLF~\cite{Burnett2019-DBS}.
In contrast, most other plungers exhibit smoother power-law-like behavior across averaging times, indicating a mixture of short-time white noise and longer-time drift processes.

The frequency-domain stability is shown via the Welch PSD estimates in Fig.~\ref{fig:two-day-evolution}(c). 
Across most QDs, the spectra exhibit approximately power-law decay between $f^{-1}$ and $f^{-2}$ over the accessible bandwidth.
For trajectories with strong drift and switching (notably $N_2$ and $N_8$), the frequency exponent is slightly lower than $-2$.

%%%%%%%%%%%%%%%%%%%%%%%%%%%%%%%%%%%%%%%%%%%%%
\begin{figure*}
    \centering
    \includegraphics[width=.95\linewidth]{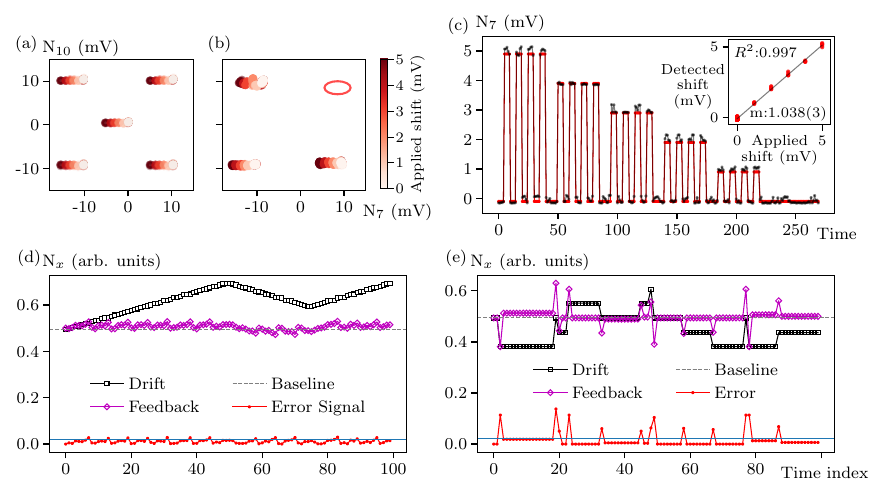}
    \caption{Detection of engineered drift on the QD array. 
    (a) Results of the charge cell center tracking when pulses of different strengths are applied to normalized plunger N$_{7}$ every 5~\si{\second}, showing the quadrilateral corners and the resulting cell centers determined through TERNS.
    (b) The cell corners identified via the MAViS interdot classifier model, with the top-left corner of the cell consistently missing.
    (c) The pulses applied to N$_{7}$ (red) and the corresponding detected change in positions of the charge cell center (black), showing perfect correlations. 
    (Inset) A linear fit to the amplitude of the detected shift vs applied shift, giving $R^2\approx 1$.
    Simulated trajectory of the center of the charge state cell with (purple) and without (black) real-time feedback applied to a map subject to (d) linear drift and (e) random jumps. 
    The error signal is shown in red with the feedback threshold and gain factor set to $(F_g, \epsilon) = (3.4, 0.02)$ respectively.
    }
    \label{fig:tracking_and_feedback}
\end{figure*}
%%%%%%%%%%%%%%%%%%%%%%%%%%%%%%%%%%%%%%%%%%%%%

While a single TLF may dominate, the underlying $1/f$ "noise floor" is generally attributed to a large ensemble of non-interacting TLFs with a wide distribution of switching rates~\cite{Dutta1981-LFF}.
By integrating the remaining noise power over a defined frequency band, we can derive an approximate, effective TLF density, $N_{\rm TLF}$. 
This metric serves as a valuable figure of merit for comparing material quality and fabrication consistency across different devices or wafers.
We report a representative noise amplitude (colored bars) at $f=100~\mu$\si{\hertz} in Fig.~\ref{fig:two-day-evolution}(d), highlighting a wide distribution in noise amplitude across the device, with N$_3$ having the highest noise amplitude.
The corresponding fitted $\alpha$ coefficients are shown with black markers.
We see that for all plungers, the exponent is lower than -1 and for some, lower than -2, suggesting that a drift process dominates the dynamics of each plunger.
A more detailed classification of microscopic noise sources is left to future work.
Related approaches for inferring charge-noise locality have been explored in Refs.~\cite{Rojas2025-ICN, Yoneda2023-NCS}.

Finally, we quantify spatial correlations by computing the pairwise Pearson correlation coefficients between drift trajectories, shown in Fig.~\ref{fig:two-day-evolution}(e).
The measurement time for all DQD pairs is around 5~\si{\second} and the time between consecutive data points is 30~\si{\minute}. 
This large-scale separation between time scales is sufficient to estimate low-frequency noise and correlations, since a 5-second measurement can be thought of as an instant relative to 30~\si{\minute}.
As expected, repeated measurements of the same plunger yield high correlations (typically $>0.8$). 
Across distinct plungers, correlations are generally stronger for nearby gates and decay with inter-plunger separation, with N$_1$, N$_{4}$, and N$_{10}$ being the least correlated with other plungers.

In Fig.~\ref{fig:two-day-evolution}(f), we plot the mean correlation versus distance, with error bars indicating the standard error in the mean, which provides an alternative visual of this trend.
An exponential fit to the data yields a correlation length of $(188\pm38)$~\si{\nano\meter}~\cite{Zajac16-SGA}. 
Similar correlation lengths of order $200$~\si{\nano\meter} have been previously reported in atom-defined QD arrays~\cite{Donnelly2025-NCA}.

%%%%%%%%%%%%%%%%%%%%%%%%%%%%%%%%%%%%%%%%%%%%%%%%%%%%%%%%%%%%%
\subsection{Real-time detection and feedback}
\label{ssec:demo-feedback}
%%%%%%%%%%%%%%%%%%%%%%%%%%%%%%%%%%%%%%%%%%%%%
A key requirement for autonomous calibration is the ability to detect abrupt electrostatic shifts and infer the corrective voltage update needed to return the device to its target operating point. 
We test the limits of our tracking and stabilization framework by introducing controlled, time-dependent perturbations to a selected DQD and quantifying how accurately the tracking system recovers the induced motion of the $(n_i,n_j)$ charge state cell.

%%%%%%%%%%%%%%%%%%%%%%%%%%%%%%%%%%%%%%%%%%%%%
\subsubsection{Engineered jumps in an experimental DQD}
\label{ssec:demo_feedback}
%%%%%%%%%%%%%%%%%%%%%%%%%%%%%%%%%%%%%%%%%%%%%
To test the tracking system, we consider the DQD defined by plungers N$_7$ and N$_{10}$ and apply a square wave to the gate N$_{7}$ every 5~\si{\second}, to induce discrete shifts of the CSD. 
By sweeping the square-wave amplitude, we generate a family of time-dependent jumps spanning the range of interest and evaluate whether the extracted charge-cell center follows these perturbations.

The results are summarized in Fig.~\ref{fig:tracking_and_feedback}.
Panel (a) shows the locations of the corners defining the quadrilateral determined using TERNS, along with the resulting cell center estimates, for square-wave amplitudes ranging from $5$~\si{\milli\volt} to $0$~\si{\milli\volt}.
Panel (b) shows the corresponding corners detected through the MAViS interdot-only approach, with the top-left interdots consistently missing, corroborating the fact that using the transition lines is more robust.

The applied barrier waveform (red) together with the corresponding inferred displacement of the $(n_i,n_j)$ cell center (black) is shown in panel (c).
The detected drifts exhibit the expected bimodal distribution corresponding to the two square-wave levels for each of the pulse intensity, and the inferred jump magnitudes agree closely with the applied perturbations (inset, $R^2=0.997$).
The time-dependent fluctuations are clearly resolved down to $\gtrsim 0.2$~\si{\milli\volt}, and the resulting center re-centering precision is on the order of $\sim 0.1$~\si{\milli\volt}, comparable to the scan resolution.
Together, these results demonstrate that using TERNS provides substantially more robust drift detection than relying solely on interdot features.

%%%%%%%%%%%%%%%%%%%%%%%%%%%%%%%%%%%%%%%%%%%%%
\begin{figure*}
    \centering
    \includegraphics[width=.95\linewidth]{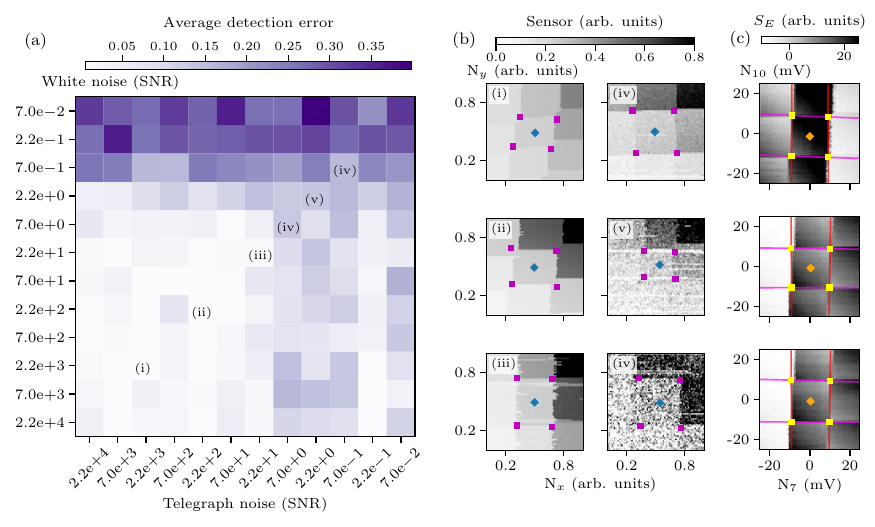}
    \caption{Performance of TERNS on simulated data. 
    (a) The distribution of the mean detected error for different levels of signal-to-noise ratio (SNR).
    For each combination of white noise and telegraph noise levels, the error is calculated across 10 realizations.
    We find that as the SNR decreases, the performance of the detection algorithm deteriorates. 
    At about white noise-induced SNR levels of $\approx 0.7$, the simulated charge stability maps become completely noisy. 
    (b) A single realization of a charge stability map with noise levels corresponding to the labels in (a), along with the predicted interdots (magenta) and the center of the central charge state (teal). 
    (b) Example experimental datasets that we study in this paper.
    }
    \label{fig:sim-benchmarking}
\end{figure*} 
%%%%%%%%%%%%%%%%%%%%%%%%%%%%%%%%%%%%%%%%%%%%%

%%%%%%%%%%%%%%%%%%%%%%%%%%%%%%%%%%%%%%%%%%%%%
\subsubsection{Simulation-based validation of the stabilization feedback}
%%%%%%%%%%%%%%%%%%%%%%%%%%%%%%%%%%%%%%%%%%%%%
To test the detection and feedback behavior under controlled linear drift and random jumps, we use the constant-capacitance model simulator \textsc{QArray}~\cite{vanStraaten24-QAR, vanStraaten24-QAC}.
We assume the plunger gates used to define the DQD have already been orthogonalized and normalized such that, in the absence of perturbations, the $(n_i,n_j)$ charge state cell is centered in the CSD.

Similar to a typical device-drift process, we simulate linear drift to cause the center of the charge-state cell to shift linearly over time. 
Under the stochastic jump process, the CSD undergoes discrete random shifts.
To implement random charge jumps, we consider the Poisson jump process, $P(\gamma)$, with mean waiting time $\gamma=5$. 

The ability to precisely track the center of the charge-state cell in simulations, while the underlying device exhibits linear drift and random jumps, enables us to apply real-time feedback and corrections. 
Figure~\ref{fig:tracking_and_feedback}(d) shows the feedback loop implemented for the linear drift, where in the absence of feedback, the plunger voltage is seen to drift with time (black circles), while with correction, the charge state is maintained at the desired reference point.
The charge state cell center trajectory under random jumps with and without feedback is shown in Fig.~\ref{fig:tracking_and_feedback}(e). 
For all our results, we set the feedback gain factor, $F_g = 3.4$.

Each iteration of the simulated feedback loop involves localizing the charge-state cell's center within the CSD, assessing its drift relative to prior measurements, and, if necessary, applying a corrective pulse.
To determine whether correction is necessary, the distance error between the current and prior center position is computed and compared against the tolerance threshold $\epsilon$. 
Throughout the tests, the tolerance is set to $\epsilon=0.02$ to strike a balance between the desired precision and the frequency of adjustments.

When the center distance error exceeds the tolerance, a proportional-only correction to the shift parameter in the simulator is applied~\cite{Bechhoefer2005-FFP}, with gain set by the feedback factor, see Eq.~(\ref{eq:gain_factor}).
This closed-loop procedure stabilizes the operating point against large shifts and large stochastic jumps while leaving small fluctuations unchanged.

%%%%%%%%%%%%%%%%%%%%%%%%%%%%%%%%%%%%%%%%%%%%%%%%%%%%%%%%%%%%%
\subsection{Testing the limits: how noisy can we get?}
\label{ssec:noisy-sim}
%%%%%%%%%%%%%%%%%%%%%%%%%%%%%%%%%%%%%%%%%%%%%
To quantify the robustness of the tracking system against degraded data quality, we benchmark its performance on increasingly more noisy synthetic CSDs.
Since the ML models in MAViS are trained on CSDs simulated using the QDFlow package~\cite{Zwolak18-QLD, Buterakos25-QDF}, we generate CSDs for testing using the \textsc{QArray} simulator~\cite{vanStraaten24-QAR, vanStraaten24-QAC}.
In simulations, we can accurately label interdot, vertical, and horizontal transitions and calculate the ground-truth charge-cell center.
Thus, these numerical experiments provide an unbiased and systematic approach to determining the noise regimes in which the reliable extraction of the $(n_i,n_j)$ charge-cell center remains possible, and to identifying failure modes relevant to automated tuning workflows.

We focus on two standard classes of noise: (i) additive white noise to model sensor/readout noise and (ii) telegraph noise to model discrete fluctuators in the device.
The white noise is modeled as the output noise relative to the Lorentzian peak encountered in the sensor data during sweeps of different plunger voltages.
Telegraph noise models internal noise in the device, wherein we calculate the potential at the sensor QD and how it is perturbed relative to the charge peaks.
For more detailed background on noise modeling, we refer the reader to ~\cite{vanStraaten24-QAR} and Appendix.~\ref{app:noise}.
We also randomly vary the latching probabilities.
For each noise setting, we evaluate ten independent realizations and the corresponding center distance error $\epsilon$ with respect to the ground-truth cell center.
To quantify the overall detection error for each noise combination, we use the average distance $D$ between the extracted and ground truth center
\begin{equation}
D = \frac{1}{N}\sum\nolimits_{i=1}^N r_c(t_i)
\end{equation}
where $N=10$ is the number of measurement repetitions and $t_i$ is used to indicate the measurement realization. 

The results are summarized in Fig.~\ref{fig:sim-benchmarking}(a).
To align with experimental data, we quantify the noise level using the signal-to-noise ratio (SNR). 
In all simulations, we define the signal to be the range (maximum $-$ minimum) of the gradient in the noiseless simulations. 

For white noise, performance decreases gradually up to the SNR levels of about $0.7$, at which point the CSD is visually dominated by noise and center tracking fails, as shown in Fig.~\ref{fig:sim-benchmarking}(d-ii). 
This holds even for relatively low levels of telegraph noise.
Telegraph noise produces a distinct failure mode: while for fixed white-noise SNR levels below $0.7$, increasing telegraph-noise strength increases the error, at high telegraph-noise SNR (around $0.22$), the inferred center can shift by one charge cell, consistent with misidentification of the target charge-state boundaries.

Overall, we find that the tracking system remains accurate across a broad range of conditions, with the average detection error below $10~\%$ for noise levels with a signal-to-noise ratio (SNR) $\geq 0.7$.
The level of noise observed in the majority of the experimental data is below these levels, see Fig.~\ref{fig:sim-benchmarking}(b) for examples of experimentally acquired CSD and Fig.~\ref{fig:sim-benchmarking}(c) for examples of simulated CSD at selected noise levels.
We also find experimental data with higher levels of noise where tracking performs well with errors $< 10 \%$.
Overall, the noise levels encountered in the experimental data fall within the regime where the tracking method maintains low error.

%%%%%%%%%%%%%%%%%%%%%%%%%%%%%%%%%%%%%%%%%%%%%%%%%%%%%%%%%%%%%
\section{Summary and outlook}
\label{sec:conclusion}
%%%%%%%%%%%%%%%%%%%%%%%%%%%%%%%%%%%%%%%%%%%%%%%%%%%%%%%%%%%%%
In this work, we present and experimentally validate TERNS---a framework for automated tracking, electrostatic recalibration, and noise spectroscopy---in a large 2D QD array. 
By exploiting the full transition-line topology of CSDs, rather than relying on isolated interdot features or single-point sensing, TERNS provides a robust, device-level sensor of electrostatic drift and charge rearrangements.
Using a 2D planar QD array in germanium, we demonstrated the system's ability to (i) track the target charge-state cell over an extended timescale, quantifying its motion in the normalized plunger space; (ii) detect and characterize engineered operating-point shifts with sub-millivolt sensitivity; (iii) perform time-domain noise spectroscopy without applying additional excitation signals; and (iv) quantify spatial correlations of electrostatic fluctuations across the device.
By combining the tracking system with a \textsc{QArray} simulator~\cite{vanStraaten24-QAR, vanStraaten24-QAC}, we showed its utility as a practical building block for autonomous calibration workflows in dense multi-QDs architectures.

Beyond enabling automation, the extracted trajectories yield direct physical insight into the QD device noise environment. 
Using Allan deviation and PSD analyses, we resolved signatures consistent with single-fluctuator-driven slow drift and quantified substantial gate-to-gate variability in noise levels across the device. 
By applying the same procedure across multiple DQD pairs, we further measured spatial correlations of the electrostatic fluctuations and extracted an average correlation length of  $(188\pm38)$~\si{\nano\meter} across the array.

Over the two-day measurement, the low-frequency spectra of the extracted fluctuations were well described by an approximate power law $\propto f^{-\alpha}$, with behavior approaching an $f^{-2}$-like regime for the most unstable gates. 
Corroborated by Allan deviation analysis, this suggests that long-term instability in our system is dominated by individual two-level fluctuators superimposed on a random-walk drift. 
These results highlight that transition-network-based tracking can simultaneously capture slow spectral components and faster switching events associated with discrete charge rearrangements and individual defects.
We note that, while the present work quantifies noise signatures through Allan/PSD metrics and spatial correlations, a detailed microscopic attribution of specific noise mechanisms and defect origins remains an important direction for future study.

A key advantage of this approach is that it provides diagnostic access to the electrostatic environment affecting qubits in the \emph{core} of dense arrays, where individual QDs may be only weakly coupled to external sensors and therefore difficult to probe using conventional readout geometries. 
This capability is directly relevant for scalable architectures in which calibration must be performed across many QDs simultaneously and repeatedly.
Finally, complementing the experimental demonstration, the simulation benchmarks establish the operational limits by quantifying the data-quality requirements for reliable tracking under controlled white and telegraph noise.

Looking forward, the stabilization performance can be further enhanced by integrating more advanced feedback protocols, including proportional-integral-derivative (PID) control, adaptive gains, and model-based controllers~\cite{Bechhoefer2005-FFP}, as well as by propagating uncertainty estimates from the extracted transition geometry to enable confidence-aware updates.
More broadly, as quantum processors scale to larger 1D and 2D arrays, long-duration operation will require continuous monitoring, rapid diagnosis of drift and charge jumps, and
automated retuning of operating points, making manual recalibration highly impractical. 
TERNS framework introduced here advances this goal by unifying (i) a robust sensing primitive for tracking charge-state geometry, (ii) quantitative noise spectroscopy and spatial correlation analysis, and (iii) a pathway toward closed-loop re-centering.
Together, these elements form a foundation for the autonomous calibration and stabilization layers that will be essential for maintaining qubit operating points and gate performance over the long times required for scalable, fault-tolerant operation.

%%%%%%%%%%%%%%%%%%%%%%%%%%%%%%%%%%%%%%%%%%%%%%%%%%%%%%%%%%%%%
\begin{acknowledgments}
This research was sponsored in part by the U.S. Army Research Office (ARO) under Awards No. W911NF-23-1-0110 and No. W911NF-23-1-0258. 
We acknowledge support from the European Union through the IGNITE project with Grant Agreement No. 101069515 and from the Dutch Research Council (NWO) via the National Growth Fund program Quantum Delta NL (Grant No. NGF.1582.22.001).

The views, conclusions, and recommendations contained in this paper are those of the authors and are not necessarily endorsed nor should they be interpreted as representing the official policies, either expressed or implied, of the U.S. Army Research Office (ARO) or the U.S. Government. 
The U.S. Government is authorized to reproduce and distribute reprints for Government purposes notwithstanding any copyright noted herein. 
Any mention of commercial products is for information only; it does not imply recommendation or endorsement by the National Institute of Standards and Technology.
\end{acknowledgments}
%%%%%%%%%%%%%%%%%%%%%%%%%%%%%%%%%%%%%%%%%%%%%%%%%%%%%%%%%%%%%

\appendix

%%%%%%%%%%%%%%%%%%%%%%%%%%%%%%%%%%%%%%%%%%%%%%%%%%%%%%%%%%%%%
\section{Algorithmic and implementation details}
\label{app:algorithms}
%%%%%%%%%%%%%%%%%%%%%%%%%%%%%%%%%%%%%%%%%%%%%%%%%%%%%%%%%%%%%
We use an ensemble of U-Net-like convolutional neural networks~\cite{ronneberger2015-UCN} to produce, for each pixel within a CSD, class probabilities for three transition types: vertical, horizontal, and interdot.
From the ensemble output, we must extract only the transition lines that bound a specific charge state of interest, which we denote $(n_i,n_j)$, rather than all lines of a given orientation.
To estimate the two vertical transitions adjacent to the $(n_i,n_j)$ charge state, we first zoom in so that the Hough-transform window tightly spans the charge-state region of interest and contains the transition endpoints. 
We refer to this procedure as a \textit{dynamic-windowed Hough transform}.

In the Hough space, we search for a pair of nearly parallel vertical lines, allowing a maximum angular variation of $1.5^\circ$ and a spacing consistent with the plunger normalization. 
In our implementation, the expected separation is $20~\si{\milli\volt}$. 
Plunger normalization enforces a common transition spacing across devices, improving the robustness of the windowed Hough search.

If two distinct peaks cannot be identified in the Hough space under the distance and angle constraints, we iteratively enlarge the window and repeat the search, up to the full image extent, selecting the best pair of nearly parallel lines that bracket the target charge state with spacing constrained to be near $20~\si{\milli\volt}$. 
As a final fallback, if only a single reliable transition line is detected, we keep that line and generate the missing partner by reflecting it at the plunger-normalized spacing.

When horizontal transitions are missed due to poor data or weak gradients, we rely solely on the vertical transitions and locate the central charge state by identifying where the vertical lines terminate at the interdot transition. 
From these, we obtain all four transition lines, their intersections (interdot transitions), and the center of the charge state cell.

%%%%%%%%%%%%%%%%%%%%%%%%%%%%%%%%%%%%%%%%%%%%%%%%%%%%%%%%%%%%%
\section{Noise in simulations}
\label{app:noise}
%%%%%%%%%%%%%%%%%%%%%%%%%%%%%%%%%%%%%%%%%%%%%%%%%%%%%%%%%%%%%
The noise models used in Sec.~\ref{ssec:demo_feedback} and Sec.~\ref{ssec:noisy-sim} introduce noise at multiple stages of the sensor-signal simulation. 
Telegraph noise is modeled as a fluctuating charge that is capacitively coupled to the sensor QD. 
The noise amplitude parameter quantifies the coupling strength: an amplitude of $1$ corresponds to the effect of a single added electron on the sensor.
The resulting sensor-dot potential, therefore, depends on the gate voltages, the charge state of the QDs, and the instantaneous fluctuator state. 
This potential is convolved with a Lorentzian to simulate Coulomb peaks, with the peak height normalized to unity. 

Additive white noise is applied to the output signal to model measurement/readout noise. 
The white-noise amplitude is defined relative to the unit peak height; in particular, an amplitude of $1$ corresponds to an SNR of $1$. 
These noise implementations are described in Ref.~\cite{vanStraaten24-QAR}.

%%%%%%%%%%%%%%%%%%%%%%%%%%%%%%%%%%%%%%%%%%%%%%%%%%%%%%%%%%%%%
%
%%%%%%%%%%%%%%%%%%%%%%%%%%%%%%%%%%%%%%%%%%%%%%%%%%%%%%%%%%%%%

%%%%%%%%%%%%%%%%%%%%%%%%%%%%%%%%%%%%%%%%%%%%%%%%%%%%%%%%%%%%%
\end{document}